\documentclass[twocolumn,floatfix,superscriptaddress,a4paper,showpacs,showkeys,nofootinbib,reprint,prc]{revtex4-1}
\usepackage{latexsym}
\usepackage{graphicx}
\usepackage[colorlinks=true,linktocpage=true,linkcolor=blue,citecolor=blue,allcolors=blue]{hyperref}
\usepackage{url}
\usepackage[utf8]{inputenc}
\usepackage{enumerate}
\usepackage{color}
\usepackage{xcolor}
\usepackage{amsmath,amssymb}
\usepackage[english]{babel}
\usepackage{hyperref}
\usepackage{url}
\usepackage{needspace}

\newcommand{\tf}{Thermal-FIST}
\begin{document}

\title{The structure of the $f_0(980)$ from system size dependent hadronic resonance ratios in p+p, p+Pb, and Pb+Pb collisions at the LHC}

\author{Tom Reichert}
\affiliation{Institut f\"{u}r Theoretische Physik, Goethe-Universit\"{a}t Frankfurt, Max-von-Laue-Str. 1, D-60438 Frankfurt am Main, Germany}
\affiliation{Frankfurt Institute for Advanced Studies (FIAS), Ruth-Moufang-Str. 1, D-60438 Frankfurt am Main, Germany}
\affiliation{Helmholtz Research Academy Hesse for FAIR (HFHF), GSI Helmholtzzentrum f\"ur Schwerionenforschung GmbH, Campus Frankfurt, Max-von-Laue-Str. 12, 60438 Frankfurt am Main, Germany }

\author{Jan Steinheimer} 
\affiliation{GSI Helmholtzzentrum f\"ur Schwerionenforschung GmbH, Planckstr. 1, D-64291 Darmstadt, Germany}
\affiliation{Frankfurt Institute for Advanced Studies (FIAS), Ruth-Moufang-Str. 1, D-60438 Frankfurt am Main, Germany}

\author{Volodymyr Vovchenko} 
\affiliation{Physics Department, University of Houston, Box 351550, Houston, TX 77204, USA}
\affiliation{Frankfurt Institute for Advanced Studies (FIAS), Ruth-Moufang-Str. 1, D-60438 Frankfurt am Main, Germany}

\author{Christoph Herold} 
\affiliation{Center of Excellence in High Energy Physics and Astrophysics, School of Physics, Suranaree University of Technology, University Avenue 111, Nakhon Ratchasima 30000, Thailand}

\author{Ayut Limphirat} 
\affiliation{Center of Excellence in High Energy Physics and Astrophysics, School of Physics, Suranaree University of Technology, University Avenue 111, Nakhon Ratchasima 30000, Thailand}

\author{Marcus~Bleicher}
\affiliation{Institut f\"{u}r Theoretische Physik, Goethe-Universit\"{a}t Frankfurt, Max-von-Laue-Str. 1, D-60438 Frankfurt am Main, Germany}
\affiliation{Helmholtz Research Academy Hesse for FAIR (HFHF), GSI Helmholtzzentrum f\"ur Schwerionenforschung GmbH, Campus Frankfurt, Max-von-Laue-Str. 12, 60438 Frankfurt am Main, Germany }
\affiliation{GSI Helmholtzzentrum f\"ur Schwerionenforschung GmbH, Planckstr. 1, D-64291 Darmstadt, Germany}

\begin{abstract}
It is shown that the hadronic phase in ultra-relativistic heavy ion collisions can be used to understand the properties of the $f_0(980)$ resonance. In particular it is shown that the centrality dependence of the $f_0(980)/\pi$ and $f_0(980)/\phi$ ratios depends strongly on the $f_0(980)\rightarrow \overline{K}+K$ branching ratio and whether the $f_0(980)$ is produced as a $\left | \overline{q}q \right \rangle$ or $\left | \overline{s}s \right \rangle$ state. These conclusions are drawn from calculations within the partial chemical equilibrium of the HRG model within Thermal-FIST as well as with the fully non-equilibrium hybrid-transport approach UrQMD. Our findings show how the hadronic phase in heavy ion collisions can be used for studies of exotic hadron properties otherwise possible only in dedicated experiments such as PANDA.
\end{abstract}

\maketitle

\section{Introduction}
The increasing number of known hadronic states \cite{ParticleDataGroup:2022pth} has led to the establishment of the quark model \cite{Gell-Mann:1964ewy} describing color-neutral hadrons as bound multi quark states. Although the internal structure of most of the known particles can be well described by the quark model as $|\bar{q}q\rangle$ (meson) or $|qqq\rangle$ (baryon) states, the internal structure of some particles, especially of hadronic resonances like the scalar $f$ mesons, remains unclear \cite{Jaffe:1976ig,Montanet:1982zi,Close:1987er,Morgan:1990fk,Tornqvist:1982yv,Tornqvist:1995ay,Morgan:1974cm,Maiani:2004uc,ExHIC:2010gcb}. Alternative internal structures which are allowed by the quark model are tetra- or penta-quark states, glueballs, molecular meson+meson states or hybrids. Most prominent among these mysterious hadrons is the $f_0(980)$ state which is difficult to describe within the conventional quark model \cite{Achasov:1987ts}. It has been proposed as a $|\Bar{u}u+\Bar{d}d\rangle$ state \cite{Chen:2003za}, a $|\bar{K}K\rangle$ state \cite{Weinstein:1990gu,Ahmed:2020kmp,Wang:2022vga}, a tetra-quark state $|\bar{q}\bar{q}qq\rangle$ \cite{Achasov:1987ts,Weinstein:1983gd,Achasov:2020aun,Zhao:2021jss}, or a quark-antiquark gluon hybrid \cite{Birse:1996nh}. See also \cite{Oller:1997ng,Oller:1998hw,Guerrero:1998ei} for advances on the description of scalar resonances in the two meson scattering states.

In addition, only few attempts at understanding the modifications of the $f_0(980)$ in dense matter exist (see e.g. \cite{Oset:2000ev}), mostly due to its unclear nature. Nevertheless, a better understanding of the $f_0(980)$ may help to shed light on the nature of (the restoration of) chiral symmetry in hot and dense matter, if it could be better understood. 

However, final scrutinizing evidence remains to be found.
Experimentally, measurements by BESIII have observed it as a decay product and investigated its mixing with the $a_0(980)$ \cite{BESIII:2012aa,BESIII:2018ozj}, which itself may contain tetra-quark contribution \cite{Bulava:2023nvw}. The study of light exotics such as the $f_0(980)$ was also identified as one of the major goals of the PANDA experiment \cite{PANDA:2021ozp}.

Here, the recent measurement of $f_0(980)$ production in elementary p+p collisions by the ALICE collaboration \cite{ALICE:2022qnb} may help to distinguish the internal structure of the $f_0(980)$. In the ALICE paper the experimental measurement in p+p collisions was compared to theoretical models based on String excitation and decay (Herwig \cite{Bahr:2008pv}) and a partonic cascade followed by quark coalescence (AMPT \cite{Lin:2004en}). While such a comparison in p+p showed to be very sensitive on the hadronization mechanism and parameters, we will follow a different approach. We will focus on the interactions of the $f_0(980)$ with the hadronic medium, created in ultrarelativistic heavy ion collisions at the LHC, to study its properties.  

Thus, we describe $f_0(980)$ production in pp and PbPb collisions at the LHC employing a) the partial chemical equilibrium hadron resonance gas model within Thermal-FIST and b) the hybrid UrQMD model to investigate $f_0(980)$ production based on different assumptions of its quark structure and decay channels. 

Here, we have to note that in both, the UrQMD model and Thermal-FIST, the $f_0$ is treated as an independent quasiparticle degree of freedom. This means that the actual quark structure does not enter explicitly in either approach but implicitly through its production mechanism (e.g. as an $s+\overline{s}$ pair in the string or by including a strangeness suppression factor in Thermal-FIST) or its decays (either to $\pi+ \pi$ or $K+ \overline{K}$). More exotic quark configurations would therefore be difficult to implement.

\section{Methods}
To showcase the role of the hadronic phase on the $f_0(980)$ production, we will compare results from two distinctively different models. 
The first is the partial chemical equilibrium (PCE) thermal model based on the Thermal-FIST package \cite{Vovchenko:2019pjl,Motornenko:2019jha}. 
Here, the multiplicities of hadrons are related to thermal equilibrium states at given temperatures, making it insensitive to specific cross-sections and microscopic dynamics. 
However, the final $f_0(980)$ multiplicity does depend on its quark structure and decay channels, as we  show below. 
On the other hand, we will compare the thermal model Thermal-FIST results to simulations with the hydro-hybrid version of UrQMD where a fluid dynamic evolution, with particlization in chemical equilibrium, is followed by a fully microscopic transport evolution. Here again, the final observable multiplicity will depend on the $f_0(980)$ structure, as well as decay and regeneration channels.

\subsection{Thermal model}

In the thermal model approach to particle production in nuclear collisions, one evaluates hadron abundances from the partition function of a hadron resonance gas~(HRG).
In the standard formulation, the primordial hadron abundances are fixed at the stage of chemical freeze-out and described by the HRG model in the grand-canonical ensemble at fixed temperature $T$, baryochemical potential $\mu_B$, and system volume $V$.
These abundances are then only modified through resonance decay feeddown.
This base formulation of the model describes well most of the measured abundances of hadrons that are stable under strong interactions at different collision energies~\cite{Andronic:2017pug}.

The present study explores yields measured at midrapidity at LHC energies, where, to large accuracy, one can take $\mu_B = 0$.
Even at the LHC, however, the standard thermal model approach requires modifications to make it applicable to (i) small systems where canonical effects are important~\cite{Vovchenko:2019kes} and (ii) yields of short-lived resonances that can be affected strongly by the hadronic phase~\cite{Knospe:2015nva}.

These modifications have been introduced to the Thermal-FIST model and used to fit ALICE data in ~\cite{Vovchenko:2019kes} and \cite{Motornenko:2019jha}. In the following we will use the results obtained in these works as thermal reference. Details on the fitting procedure and more discussion can be found in the corresponding publications. However, for better context we will summarize the main extensions in the following paragraphs.

\subsubsection{Canonical suppression}

The canonical effects are considered in the framework of the canonical statistical model~(CSM)~\cite{Vovchenko:2018fiy}, which enforces the exact conservation of baryon number, electric charge, and strangeness across a correlation volume $V_c$.
In Ref.~\cite{Vovchenko:2019pjl}, the CSM was applied to analyze the abundances of stable hadrons in p+p, p+Pb, and Pb+Pb collisions at LHC energies and the dependence of parameters on charged multiplicity $\langle dN_{\rm ch} / d \eta \rangle$ was established.
We will utilize the parametrization from Ref.~\cite{Vovchenko:2019pjl} in our analysis.

The multiplicity dependence of the chemical freeze-out temperature $T_{\rm ch}$ reads,
\begin{align}
T_{\rm ch} = T_0 - \Delta T \ln\left(\langle dN_{\rm ch} / d \eta \rangle \right),
\end{align}
where $T_0 = 0.176$~GeV and $\Delta T = 0.0026$~GeV.
This dependence is shown in Fig.~\ref{fig:0} by the solid blue line.

The correlation volume $V_c$ corresponds to three units of rapidity for all $\langle dN_{\rm ch} / d \eta \rangle$, i.e. $V_c = 3 dV/dy$.
The volume per rapidity $dV/dy$ itself is a linear function of multiplicity, $dV/dy = v \langle dN_{\rm ch} / d \eta \rangle$, where $v = 2.4$~fm$^3$.

Some yields, particularly the $\phi/\pi$ ratio, indicate incomplete strangeness equilibration in small systems.
The strangeness saturation parameter $\gamma_S$ can be introduced for this reason. Its multiplicity dependence from Ref.~\cite{Vovchenko:2019pjl} reads
\begin{align}
\label{eq:gammaS}
\gamma_S = 1 - A \exp \left[ - \frac{\langle dN_{\rm ch} / d \eta \rangle}{B}\right],
\end{align}
with $A = 0.25$ and $B = 59$. In the present study, we optionally include the strangeness non-equilibrium effects through Eq.~\eqref{eq:gammaS}.

The canonical suppression and strangeness under-saturation effects are mainly relevant in small systems, $\langle dN_{\rm ch} / d \eta \rangle \lesssim 200$.

\subsubsection{Partial chemical equilibrium}

The chemical equilibrium thermal model neglects hadronic phase dynamics.
For this reason, it is not well suited for describing the yields of short-lived resonances such as $K^*(892)$, $\rho(770)$, and possibly $f_0(980)$, since these can be modified appreciably in the hadronic phase as indicated e.g. with UrQMD~\cite{Knospe:2015nva} or SMASH~\cite{Oliinychenko:2021enj} afterburner simulations.

This shortcoming is addressed by implementing the concept of partial chemical equilibrium~(PCE) in the hadronic phase~\cite{Bebie:1991ij,Huovinen:2007xh}.
The PCE is applied at temperatures below $T_{\rm ch}$, where one assumes that all inelastic hadronic reactions are forbidden but elastic~(e.g. $\pi\pi \leftrightarrow \pi\pi$) as well as pseudo-elastic reactions involving short-lived resonances~(e.g. $\pi\pi \leftrightarrow \rho$, $\pi K \leftrightarrow K^*$, and $\pi N \leftrightarrow \Delta$) are permitted and proceed in relative chemical and kinetic equilibrium.
The total abundances\footnote{The total abundance includes primordial~(thermal) abundances and the decay feeddown from short-lived resonances.} of hadrons that are stable on the time scales of the hadronic phase are conserved by construction\footnote{Extensions of PCE exist that also implement baryon annihilation, which does change the abundances of pion and nucleons in the hadronic phase~\cite{Vovchenko:2022xil}.}.
This means that the abundances of stable hadrons such as $\pi$, $K$, or p, are fixed at $T = T_{\rm ch}$ in line with the standard chemical equilibrium thermal model.
To maintain the conservation of stable hadron yields at $T<T_{\rm ch}$, one introduces independent effective chemical potentials~(fugacities) for each stable hadron.
The chemical potentials of short-lived resonances are determined through the effective chemical potentials of their stable decay products, e.g., $\mu_\rho = 2 \mu_\pi$, $\mu_\Delta = \mu_N + \mu_\pi$, and so on.

\begin{figure}[t]
  \centering
  \includegraphics[width=0.5\textwidth]{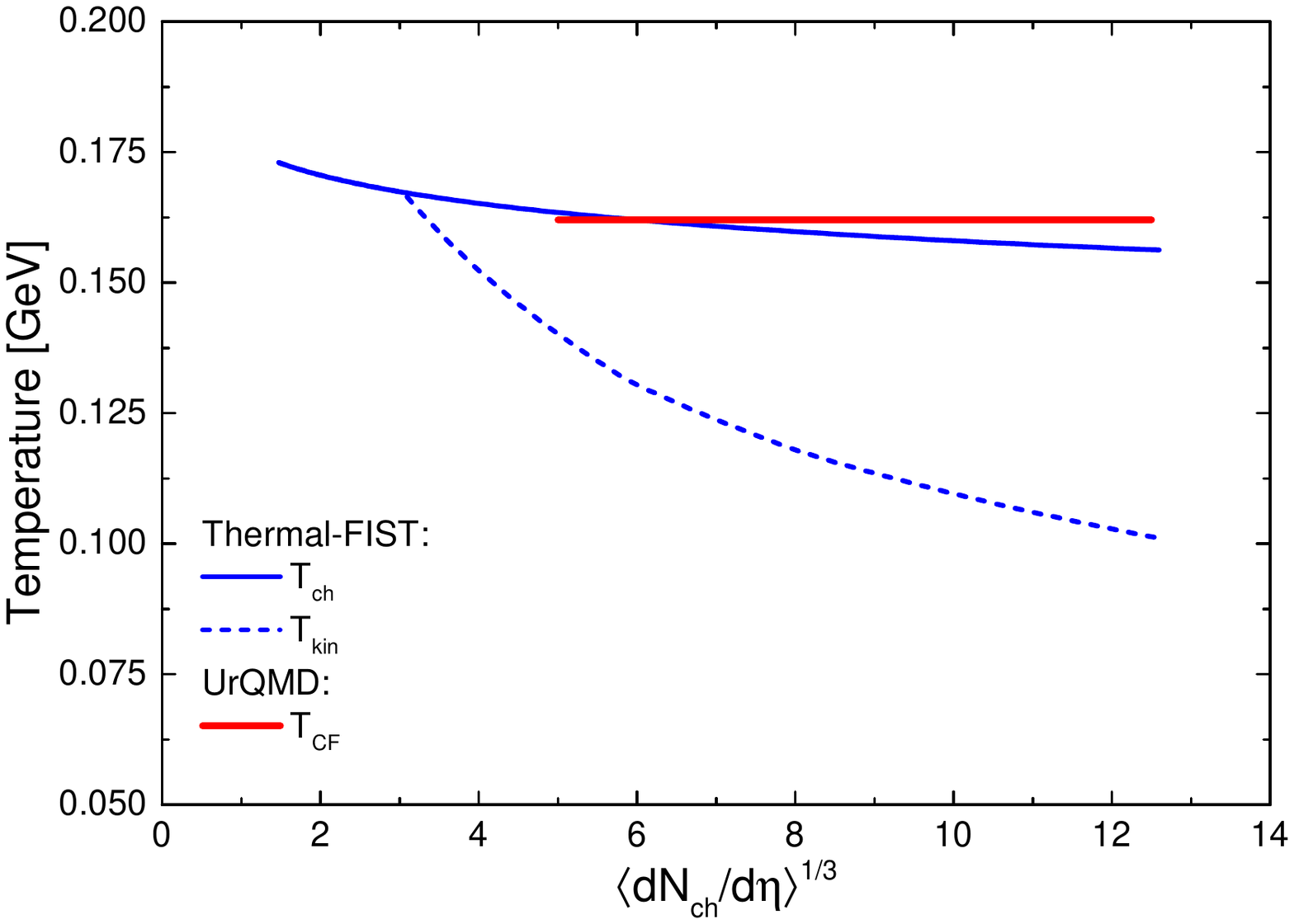}
  \caption{Parameterized chemical~(solid blue line) and kinetic freeze-out~(dashed blue line) temperatures providing the best fits to Pb+Pb data at $\sqrt{s_\mathrm{NN}}=2.76$ TeV~\cite{Vovchenko:2019pjl,Motornenko:2019jha} and used in the partial chemical equilibrium thermal model~(Thermal-FIST) in this work. The particlization (Cooper-Frye) temperature of the hybrid UrQMD simulation is shown as a red line.}
  \label{fig:0}
\end{figure}

In contrast to stable hadron yields, the yields of short-lived resonances depend on the value of temperature in the hadronic phase.
In Ref.~\cite{Motornenko:2019jha}, the yields of $K^{*0}$ and $\rho^0$ were fitted to establish the kinetic freeze-out temperature $T_{\rm kin}$ that characterizes the endpoint of hadronic phase in various systems at LHC, where it was found to be a decreasing function of charged multiplicity.
Here we use the $T_{\rm kin}$ values for different multiplicities from Table I in~\cite{Motornenko:2019jha} and perform linear interpolation in $\ln \langle dN_{\rm ch} / d \eta \rangle$ to obtain a smooth $\langle dN_{\rm ch} / d \eta \rangle$ dependence of this parameter. This dependence is shown by the dashed blue line in Fig.~\ref{fig:0}.

\subsubsection{Other implementation details}

We use an open-source package \tf~(version 1.4.2)~\cite{Vovchenko:2019pjl,TFgithub} in all our thermal model-based calculations. The charged multiplicity dependence of $T_{\rm ch}$, $T_{\rm kin}$, $V_c$, and, where applicable, $\gamma_S$, is described in the previous subsections.
One should note that \tf~presently implements PCE in the grand-canonical ensemble only, thus, the simultaneous  PCE+CSM calculation proceeds in two steps.
First, we calculate hadron abundances within the chemical equilibrium CSM. 
Second, we use the PCE model in the grand canonical ensemble to calculate the correction factors for the yields of short-lived resonances due to hadronic phase. The abundances computed in step one are then multiplied by these factors.
In practice, the PCE effects are mainly relevant in large systems, $\langle dN_{\rm ch} / d \eta \rangle \gtrsim 100$~\cite{Motornenko:2019jha}, while canonical suppression becomes relevant in small systems, $\langle dN_{\rm ch} / d \eta \rangle \lesssim 100$~\cite{Vovchenko:2019pjl} for hadrons under consideration in the present work. 
Thus, the canonical and PCE effects can be largely considered independently.

We utilize the default PDG2020 hadron list from \tf.
Two opposite scenarios concerning the quark content of $f_0(980)$ are taken: the default light quark $|\Bar{q}q\rangle$ state decaying into two pions, $f_0 \to \pi \pi$, and a $|\Bar{s}s\rangle$ state decaying into a pair of kaons, $f_0 \to \overline{K} K$.
We also implement finite resonance widths for all short-lived hadrons through energy-dependent Breit-Wigner scheme~\cite{Vovchenko:2018fmh}.
At the same time, we neglect excluded volume and baryon annihilation effects, which do not appear to have a major influence on the observables studied in this work.

\begin{figure*}[t]
  \centering
  \includegraphics[width=\textwidth]{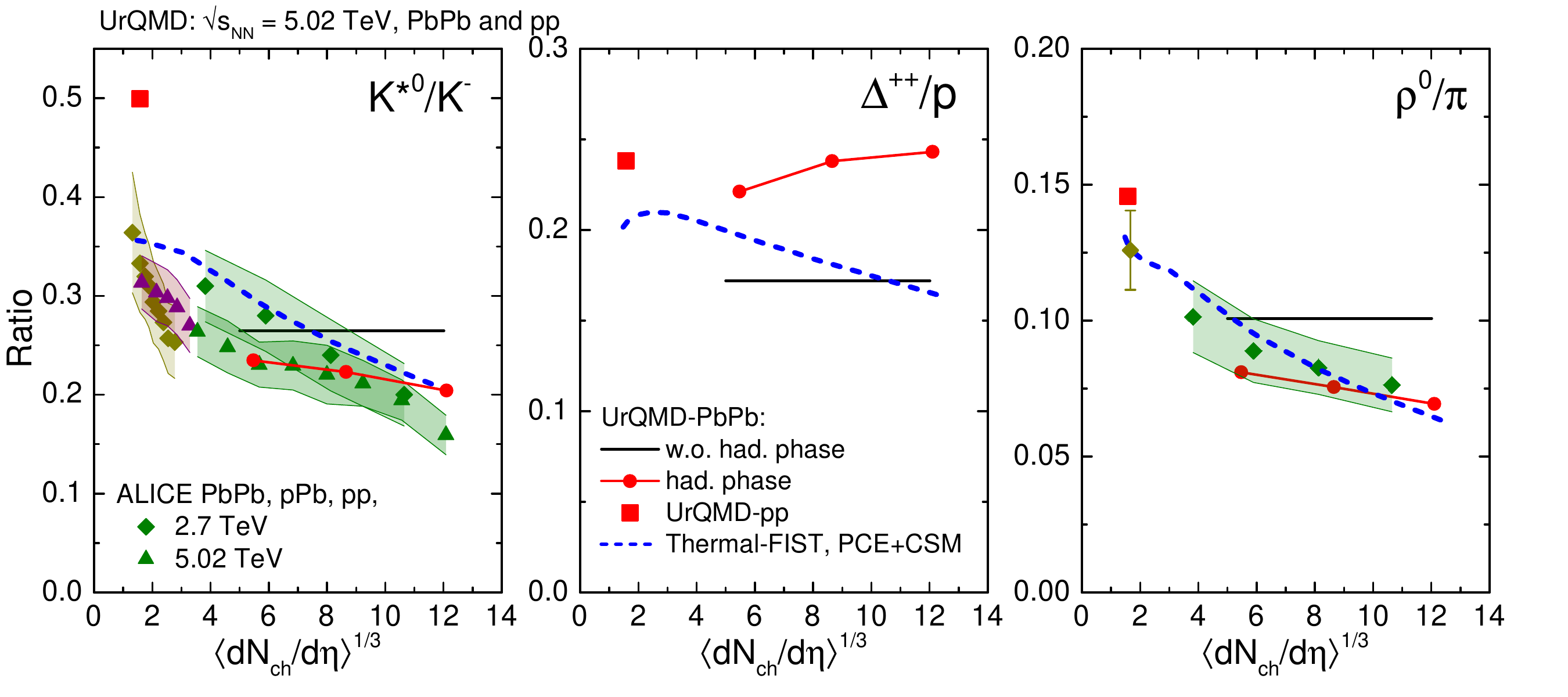}
  \caption{(Color online) The centrality dependence of the $K^{*0}/K^-$ (left panel), $\Delta^{++}/p$ (middle panel) and $\rho^0/\pi$ (right panel) ratios compared to ALICE data from different system sizes. The hybrid UrQMD calculations are shown for Pb+Pb collisions, at $\sqrt{s_{NN}}=5.02$ TeV, with a hadronic phase (full red circles) and without hadronic phase (solid black line) and for p+p collisions (full red squares). The Thermal-FIST calculations with PCE+CSM are shown as a dashed blue line. The experimental data for pp (dark yellow), pPb (purple) and PbPb (green) collisions at $\sqrt{s_{NN}}=5.02$ (triangles) and $2.7$ TeV (diamonds) are also shown \cite{ALICE:2014jbq,ALICE:2021xyh,ALICE:2023ifn,ALICE:2021ptz,ALICE:2018qdv,ALICE:2016sak,ALICE:2019etb}.}
  \label{fig:resonance_ratios}
\end{figure*}

 \subsection{The UrQMD hybrid model}
For a dynamical description of the production and decay of resonances, as well as rescattering of the decay products, fully dynamical simulations are necessary. We will therefore employ the Ultra-relativistic Quantum Molecular Dynamics \cite{Bass:1998ca,Bleicher:1999xi} (UrQMD) transport model to calculate the evolution of hadrons in the final stages of the nuclear collisions. At larger center-of-mass energies a so-called hybrid-model was established \cite{Petersen:2008dd} in which the hot and dense phase of the time evolution is described by an ideal fluid dynamical simulation \cite{Rischke:1995ir}. The very early non-equilibrium stage of the collisions is modeled by multiple N+N scatterings simulated by the excitation and de-excitation of color-strings according to the string model \cite{Andersson:1983ia} (to calculate the $f_0(980)$ PYTHIA has been turned off and the native UrQMD string based on the LUND model has been used) which is also used in the description of elementary p+p collisions at the LHC (including hadronic rescattering in UrQMD).

The transition from the fluid description, back to the transport description occurs on an iso-energy-density hypersurface corresponding to a temperature of approximately 162 MeV at vanishing net baryon density. This temperature is indicated as red line in Fig.~\ref{fig:0}. This hypersurface is then used to sample hadrons (including a full set of hadronic resonances) according to the Cooper-Frye prescription \cite{Cooper:1974mv,Huovinen:2012is} assuming equilibrium distributions in the respective cell\footnote{Note, that at this point the width of the resonance is neglected and only peak masses are used.}. The created particles then further interact in the cascade part of the UrQMD model. All kinetic reactions cease dynamically based on the local interplay of the expansion rate and the scattering rate. The equation of state used in the hydrodynamic evolution contains a smooth crossover between a hadron resonance gas and a deconfined quark-gluon-plasma \cite{Motornenko:2019arp}.

\subsection{Known resonance ratios}

To benchmark the hybrid UrQMD model and the Thermal-FIST model with PCE+CSM, and validate their ability to describe other resonance multiplicities, we first calculate the centrality dependence of well-studied resonance ratios, i.e. the $K^{*0}(892)/K^-$, $\Delta^{++}(1232)/p$ and $\rho^0(770)/\pi$ ratios, as function of system size. For the sake of brevity we will drop the round brackets depicting the mass number. Fig. \ref{fig:resonance_ratios} shows the centrality dependence of the $K^{*0}/K^-$ (left panel), $\Delta^{++}/p$ (middle panel) and $\rho^0/\pi$ (right panel) ratios compared to ALICE data \cite{ALICE:2014jbq,ALICE:2021xyh,ALICE:2023ifn,ALICE:2021ptz,ALICE:2018qdv,ALICE:2016sak,ALICE:2019etb}. The hybrid UrQMD calculations are shown for Pb+Pb collisions with hadronic phase (full red circles) and without hadronic phase (solid black line), for p+p collisions (full red squares) while the Thermal-FIST calculations with PCE+CSM are shown as a dashed blue line and the experimental data is depicted as green symbols.

It can be observed that the $K^{*0}/K^-$ and the $\rho^0/\pi$ ratios show a strong suppression in the experimental data from elementary to central collisions due to rescattering of the decay daughters in the dense medium. This behavior is well reproduced in the canonical statistical model with partial chemical equilibrium \cite{Motornenko:2019jha}, with the dominant effect coming from PCE. 
The hybrid UrQMD calculations without hadronic afterburner resemble the grand canonical ratios at fixed energy density. 
The lack of the rescattering phase therefore provides a flat result as we do not use a system size dependent particlization temperature for the hybrid model. 
The full hybrid UrQMD simulation with hadronic afterburner however reproduces the trend of the suppression in the central to mid-peripheral collisions and also continues the trend to p+p collisions (also found in previous studies \cite{Knospe:2015nva,Knospe:2021jgt}). 

The two model calculations show qualitatively different results for the $\Delta^{++}/p$ ratio. The CSM+PCE calculations within Thermal-FIST show a suppression of the ratio towards central collisions.

This suppression results from two effects: the suppression of $\Delta^{++}/p$ in the hadronic phase due to the drop in kinetic freeze-out temperature and, to a smaller extent, the multiplicity dependence of chemical freeze-out temperature.
The hybrid UrQMD calculations with hadronic afterburner first show an enhancement in p+p, compared to peripheral Pb+Pb collisions, but an increase of the ratio towards central Pb+Pb collisions. The enhancement in p+p can be explained by the selection of resonances in the string routine which tends to favor the lowest excitable states, i.e. the Delta. The additional enhancement of the $\Delta^{++}/p$ from peripheral to central Pb+Pb collisions has a different reason. Here, it is worthwhile to note that the pion+proton inelastic cross section is the largest of all possible hadronic reaction channels and the number of Delta excitations exceeds any other possible hadronic reaction except pion+pion scatterings. Therefore, the excitation of $\Delta$'s is the last reaction to cease \cite{Steinheimer:2017vju}. At this time the density is very low and the chance of the decay pion to rescatter with anything else but another nucleon, forming another $\Delta$, is very low and thus, the chance of reconstructing the $\Delta$ is very high, even if it is not from the last generation of decays. As a consequence, the observed $\Delta$'s in the dynamical approach appear larger than in the PCE fit. 

\begin{figure}[t]
  \centering
  \includegraphics[width=0.5\textwidth]{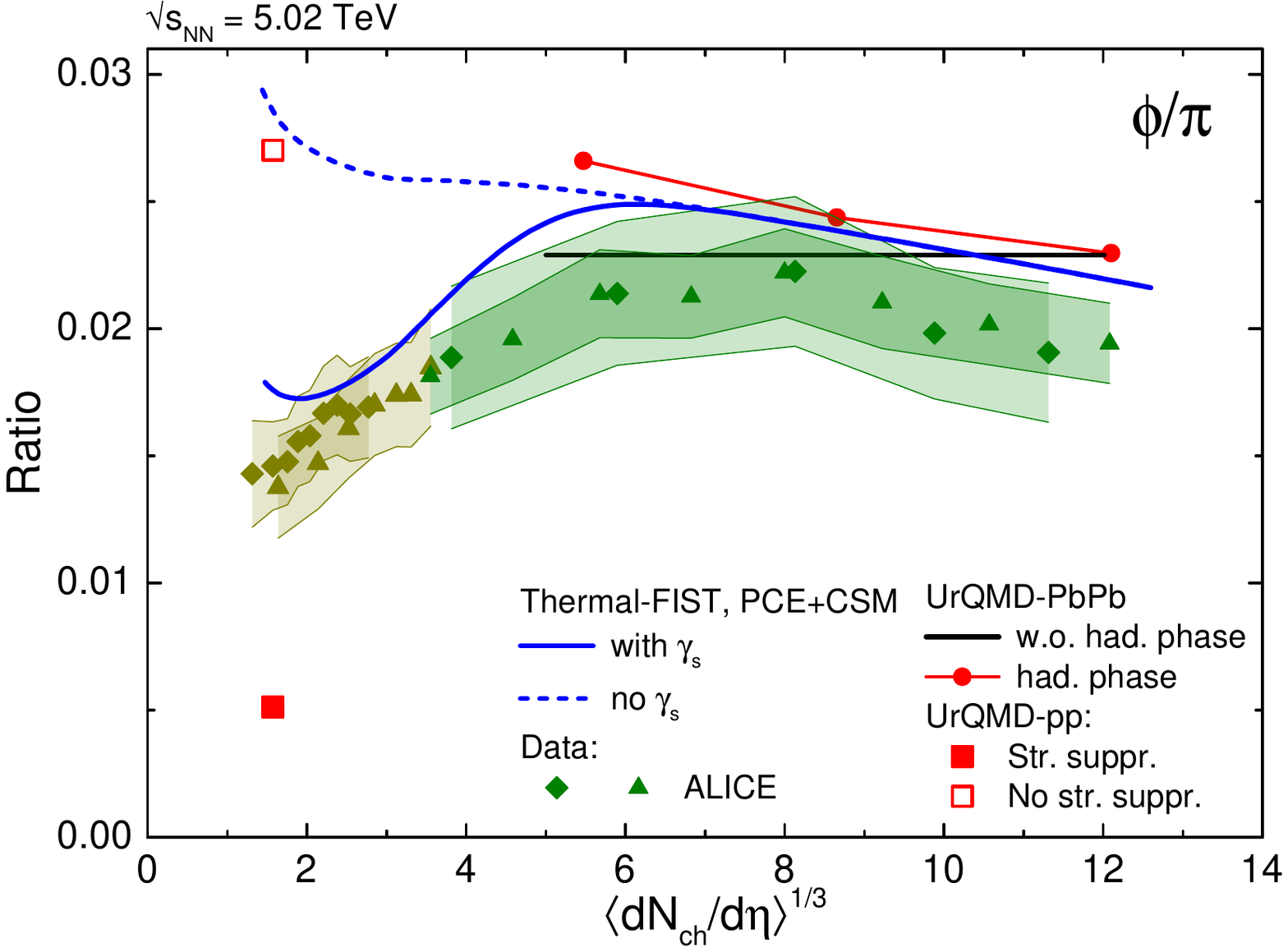}
  \caption{(Color online) The centrality dependence of the $\phi/\pi$ ratio compared to ALICE data \cite{ALICE:2021ptz,ALICE:2014jbq,ALICE:2016sak,ALICE:2019etb}. The hybrid UrQMD calculations for Pb+Pb collisions with hadronic phase are shown as full red circles and without hadronic phase as solid black line, for p+p collisions the UrQMD calculations with strangeness suppression in the string are shown as full red squares and without strangeness suppression as open red squares while the Thermal-FIST calculations with PCE+CSM and with $\gamma_s$ factor are shown as solid blue lines and without $\gamma_s$ as dashed blue lines and the experimental data is depicted by green symbols.}
  \label{fig:phi_to_pi}
\end{figure}

Of particular interest is also the $\phi/\pi$ ratio, as the $\phi$ meson is a $\overline{s}s$-state. Fig. \ref{fig:phi_to_pi} shows the centrality dependence of the $\phi/\pi$ ratio compared to ALICE data \cite{ALICE:2021ptz,ALICE:2014jbq}. The hybrid UrQMD calculations for Pb+Pb collisions with hadronic phase are shown as full red circles and without hadronic phase as solid black line, for p+p collisions the UrQMD calculations with strangeness suppression in the string are shown as full red squares and without strangeness suppression as open red squares. The Thermal-FIST calculations with PCE+CSM are shown with a $\gamma_s$ factor as solid blue lines and without $\gamma_s$ as dashed blue lines and the experimental data is depicted by green symbols. The $\gamma_s$ factor is introduced to incorporate the strangeness suppression observed for small systems and acts on each the strange and anti-strange quark in the $\phi$.

One observes several interesting features. Firstly, for heavy-ion collisions, one has $\gamma_S \simeq 1$ thus thermal results reflect chemical equilibrium, and they agree nicely with the results from the full UrQMD-hybrid simulations. Only a small suppression of $\phi$-mesons towards central collisions is observed due to the long lifetime of the $\phi$. The small increase in peripheral collisions in the hybrid model is due to few regenerations of the $\phi$ early in the short hadronic phase. The very peripheral or elementary p+p systems behave differently. Here, the strangeness suppression is an essential feature. While in the Thermal-FIST, the fit with $\gamma_s$ nicely describes the data, the UrQMD p+p results with strangeness suppression in the string even underestimates the $\phi$. This is a known feature of the string as the production of an $\overline{s} s$-state requires the creation of two pairs of $\overline{s} s$ next to each other. Both the Thermal-FIST and UrQMD results without strangeness suppression agree very well for p+p systems.

With these results we can conclude that both setups are rather well equipped to describe various hadron resonance ratios including such that involve strangeness. Differences are observed in the treatment of $\overline{s} s$-states for very small systems which will be important in the discussion of the $f_0(980)$.

\section{Results}
Before discussing the centrality dependent ratios of the $f_0(980)$ from the different models, it is first prudent to discuss its reconstruction. In the experiment, resonances are typically measured via reconstruction of their invariant mass in their respective hadronic or leptonic decay channels.

\begin{figure}[t]
  \centering
  \includegraphics[width=0.5\textwidth]{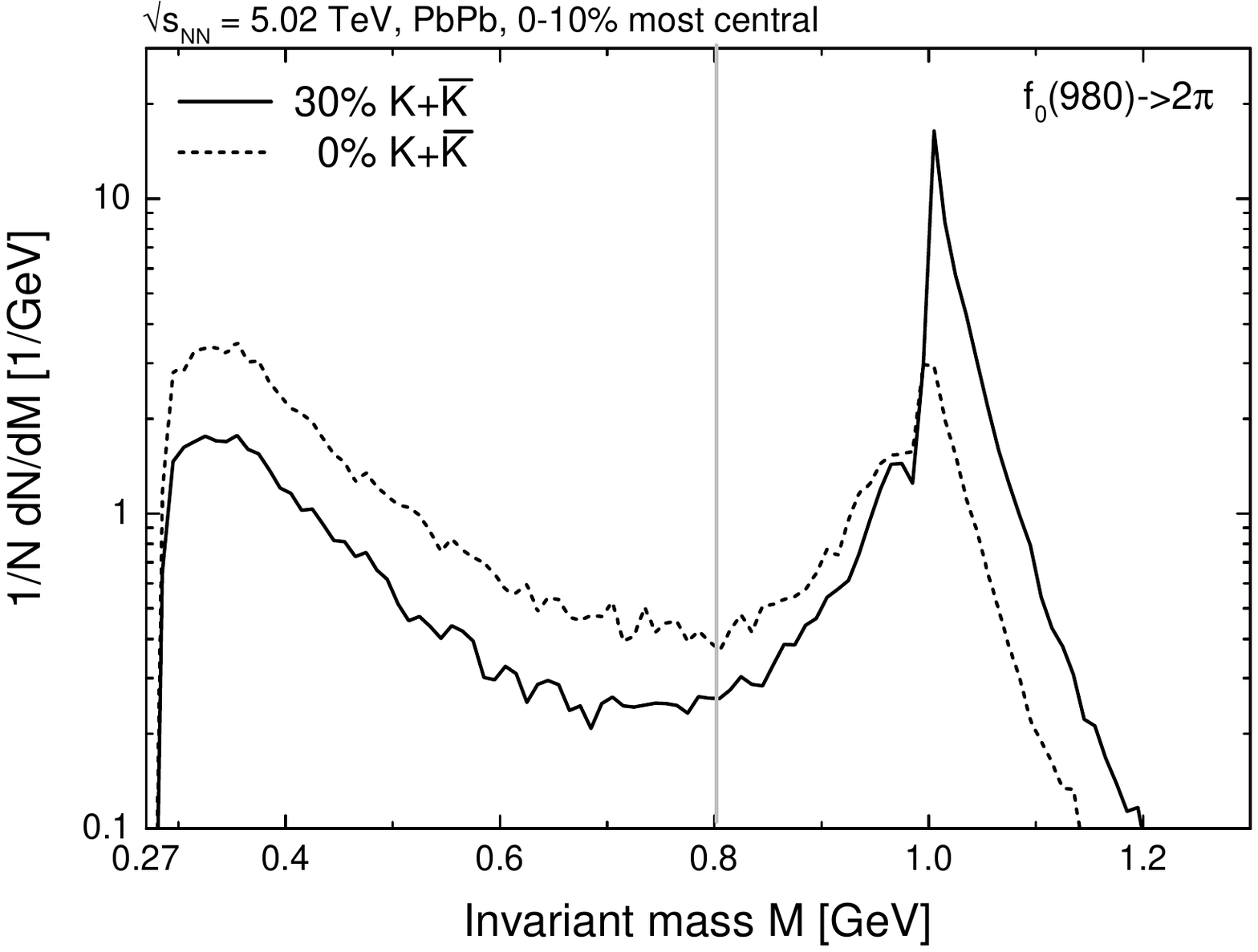}
  \caption{(Color online) Invariant mass distribution of the reconstructable $f_0(980)$ resonances reconstructed in the pion channel from central Pb+Pb collisions at $\sqrt{s_\mathrm{NN}}=5.02$ TeV from UrQMD. The solid line shows the invariant mass distributions for UrQMD calculations with a 30\% branching ratio for the kaon decay branch of the $f_0(980)$ while the calculations shown by the dashed line assume a 0\% branching ratio to the kaon channel. The vertical line shows the lower reconstruction limit reported by ALICE.}
  \label{fig:f0_mass}
\end{figure}

Although in principle a mixed event invariant mass analysis from our simulated data could be employed as well, it requires a substantial amount of statistics and is error prone due to different fits to the background shape, acceptance problems that may result in artificial structures, different efficiencies, etc. \cite{Drijard:1984pe}. In the following we will employ a simpler method and take advantage of the fact that in the simulation one can directly check which decaying resonances can be detected without signal loss, i.e. without significant rescattering of the daughter particles leading to non-observability in the invariant mass distribution \cite{Vogel:2010pb,Reichert:2019lny,Reichert:2022uha}.

We will focus only on the hadronic decay (specifically on the $\pi^+ \pi^-$ channel) of the $f_0(980)$. For other resonances, like the $\rho$, the leptonic decay is also of importance as it allows to study the spectral functions of resonances and to measure their in-medium modifications, like mass shifts or broadening of the spectral function. A very prominent example is the $\rho$ meson (Brown-Rho-Scaling \cite{Brown:1991kk}) which has been extensively studied e.g. at the NA60 experiment \cite{Damjanovic:2006bd,Damjanovic:2007qm,Arnaldi:2006jq}. In Refs. \cite{Rapp:2010sj,Rapp:2004zh} it was shown that at this energy chiral symmetry and in-medium effects mainly produce a broadening of the $\rho$ spectral function without a mass shift. The measurement of in-medium effects is competing with kinetic variations of a resonances spectral function \cite{Reichert:2022uha}. The expansion and respective cooling of the system from chemical to kinetic decoupling leads to a favored population of the low mass tail of a resonances spectral functions if it receives significant regeneration. At low energies this has been observed by FOPI \cite{Hong:1997ka,Eskef:2001qg} and recently by HADES \cite{Adamczewski-Musch:2020edy} for the $\Delta$ resonance and has been interpreted as a kinetic effect \cite{Reichert:2019lny}.

For our current study, there are several reasons why we only discuss the hadronic decay channel. The most important one is that the whole conclusion of our work relies on the effect of the hadronic rescattering phase of the collision and how it affects the measurable $f_0$. Those $f_0$ which decay in the di-electron channel are lost for any rescattering and thus carry different information.  In addition, the electromagnetic branching fraction of the $f_0$ is not well known and would simply introduce more uncertainty. Finally, the measurement of the $f_0$ in the di-lepton channel is very challenging due to the background of other vector meson decays as well as q-qbar contributions.

As we will see, also for the hadronic decay channel of the $f_0(980)$, the hadronic rescattering has a significant effect on the observed spectral shape.

\begin{figure*}[t]
  \centering
  \includegraphics[width=0.9\textwidth]{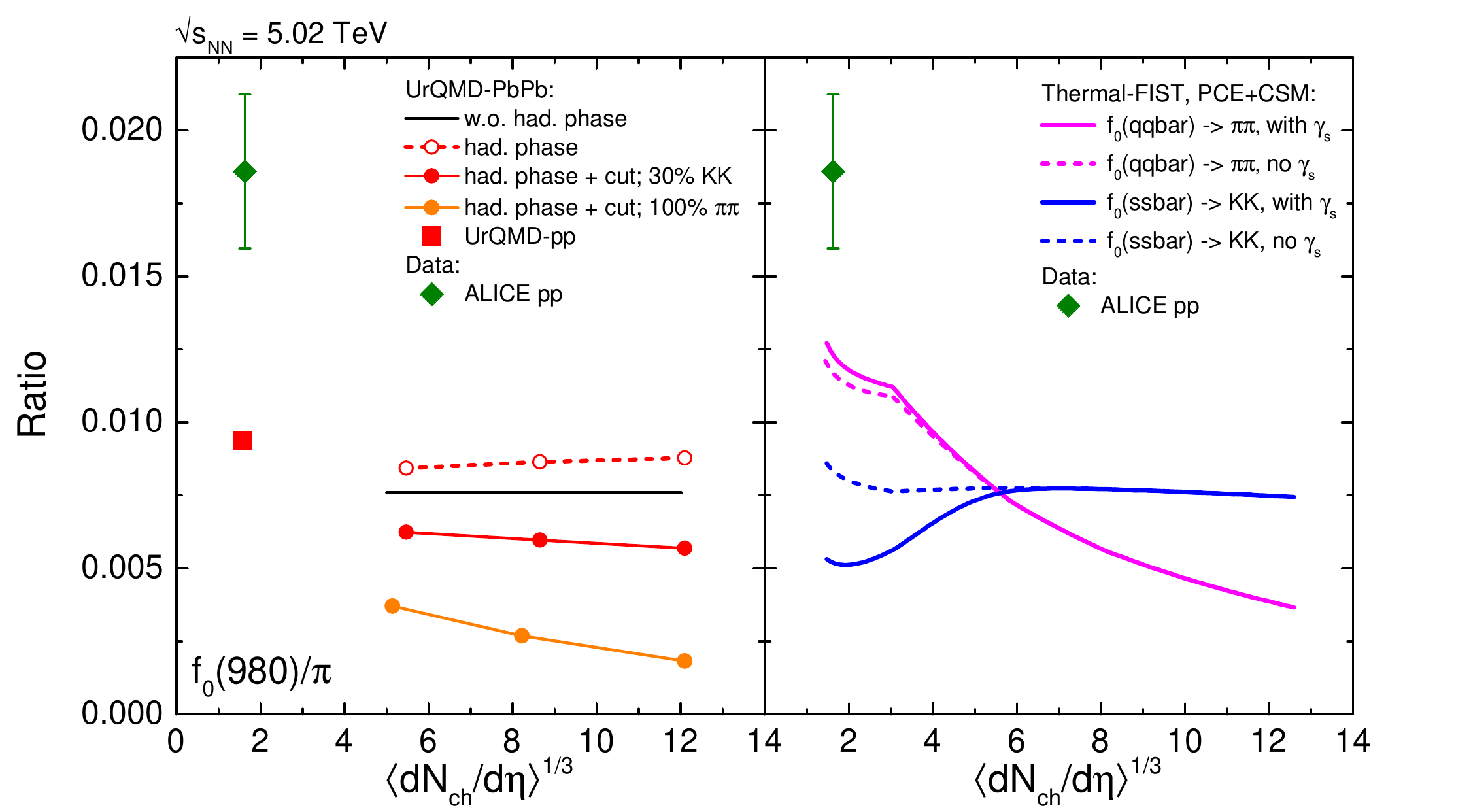}
  \caption{(Color online) The centrality dependence of the $f_0(980)/\pi$ ratio compared to ALICE data \cite{ALICE:2022qnb}. The left panel compares the UrQMD calculations to the data while in the right panel the Thermal-FIST calculations are compared to the data. The UrQMD calculations are shown for p+p collisions (full red squares) and for Pb+Pb collisions in hybrid mode without hadronic phase (solid black line) and with hadronic phase (open red circles, dashed line) with an additional phase space cut at 0.8 GeV with a 30\% branching ratio into kaons (full red circles) and at 0.8 GeV with a 100\% branching ratio into pions (full orange circles). The Thermal-FIST PCE+CSM calculations are shown considering the $f_0(980)$ as an $\langle\Bar{q}q\rangle$ state (magenta lines) with $\gamma_s$ factor (solid line) and without $\gamma_s$ factor (dashed line) and considering the $f_0(980)$ as an $\langle\Bar{s}s\rangle$ state (blue lines) with $\gamma_s$ factor (solid line) and without $\gamma_s$ factor (dashed line). Experimental data is shown by green symbols.}
  \label{fig:f0_to_pi}
\end{figure*}

ALICE has measured the $f_0(980)$ recently in the invariant mass distribution of $\pi^+\pi^-$ pairs in the invariant mass range from $0.8\leq M_{\pi^+\pi^-} \leq 1.6$ GeV \cite{ALICE:2022qnb}. The cut in the invariant mass range for the experimental reconstruction is necessary to reduce the background of other resonances including the other $f_0$ states as well as the $\rho$ meson. Since our simulation allows us to exclude such background we have the full mass range available. The 2023 PDG report \cite{Workman:2022ynf} still quotes the $f_0(980)$ decay channels into $\pi\pi$ and $\overline{K}K$ as ``seen'' leaving the branching ratios unknown as well\footnote{Note, that ALICE assumes a $30 \%$ branching into $\overline{K}K$.}. 
This leaves us the freedom to explore dependences of the observable multiplicity on the branching ratio into kaons. To do so, we employ two different branching ratios: a) a 0\% branching ratio into kaons (only pion decay) and b) a 30\% branching ratio into kaons and reduced branching to pions.
\footnote{In principle the branching ratio to the $\overline{K}K$ channel may even be larger than $30\% $ \cite{Ahmed:2020kmp,Wang:2022vga}. However, as we will see, in that case the results would simply resemble those without a hadronic rescattering, i.e. already at $30\%$ the effect of the rescattering is very minor. This is also seen in the Thermal-FIST calculation which either employs a $100\%$ pion decay OR a $100\%$ $\overline{K}K$ decay channel. To determine whether the $\overline{K}K$ is larger than $30\%$ would therefore require a very precise measurement of the centrality dependence.}

Fig. \ref{fig:f0_mass} shows the invariant mass distribution of the reconstructable $f_0(980)$ resonances reconstructed in the pion channel from central Pb+Pb collisions at $\sqrt{s_\mathrm{NN}}=5.02$ TeV from UrQMD. The solid line shows the invariant mass distributions for UrQMD calculations with a 30\% branching ratio for the kaon decay branch of the $f_0(980)$ while the calculations shown by the dashed line assume a 0\% branching ratio to the kaon channel. The vertical line shows the lower reconstruction limit reported by ALICE.

A clear observation is that the full $f_0(980)$ mass distribution shows not only the expected peak at its pole mass but also exhibits a clear maximum at very low mass, below the reconstruction threshold set by ALICE. 
Although the Breit-Wigner peak is clearly visible around the pole mass, the low mass tail exhibits a thermal behavior. This thermal tail is a result of the regeneration through pion scattering and can only appear far away from the pole\footnote{This reminds of the $\rho$ spectral function measured in $\pi\pi$ \cite{STAR:2003vqj} versus measurement in the di-leptons \cite{STAR:2013pwb,STAR:2015tnn}, where the $\rho$ with masses below $2m_\pi$ also shows a thermal tail.}. 
In order to extract the full $f_0(980)$ yield one would have to know the actual time evolution of the medium temperature and use it to fit a Breit-Wigner multiplied by the thermal weight, i.e. a phase space factor, to the extracted mass distribution. As this is not possible, the actual medium temperature is not known a priori, one could attempt to reconstruct the $f_0(980)$ in a decay channel which has almost no background from other resonances like the $\pi^0\pi^0$ channel\footnote{Vector mesons do not decay into two neutral pions.}. 

Such a measurement is of interest as one also observes a large difference in the regeneration ability of the $f_0(980)$, at low masses, when comparing the invariant mass distribution with the two employed branching ratios. Increasing the branching ratio of the decay into kaons leads to stronger regeneration close to the pole mass, and the two kaon threshold, while a smaller branching ratio into the strange channel leads to a favored regeneration of lower masses due to the larger 2-pion cross section. When measuring the full invariant mass range, one can therefore extract and pin down the branching ratio of $f_0(980)\rightarrow \bar{K}K$ more precisely as before.

As this is a challenging experimental task, for the following we will also employ the lower mass limit which ALICE uses for the extraction of their $f_0(980)$ yield by fitting a relativistic Breit-Wigner to their signal \cite{ALICE:2022qnb} and compare with the full yield.

\begin{figure*}[t]
  \centering
  \includegraphics[width=0.9\textwidth]{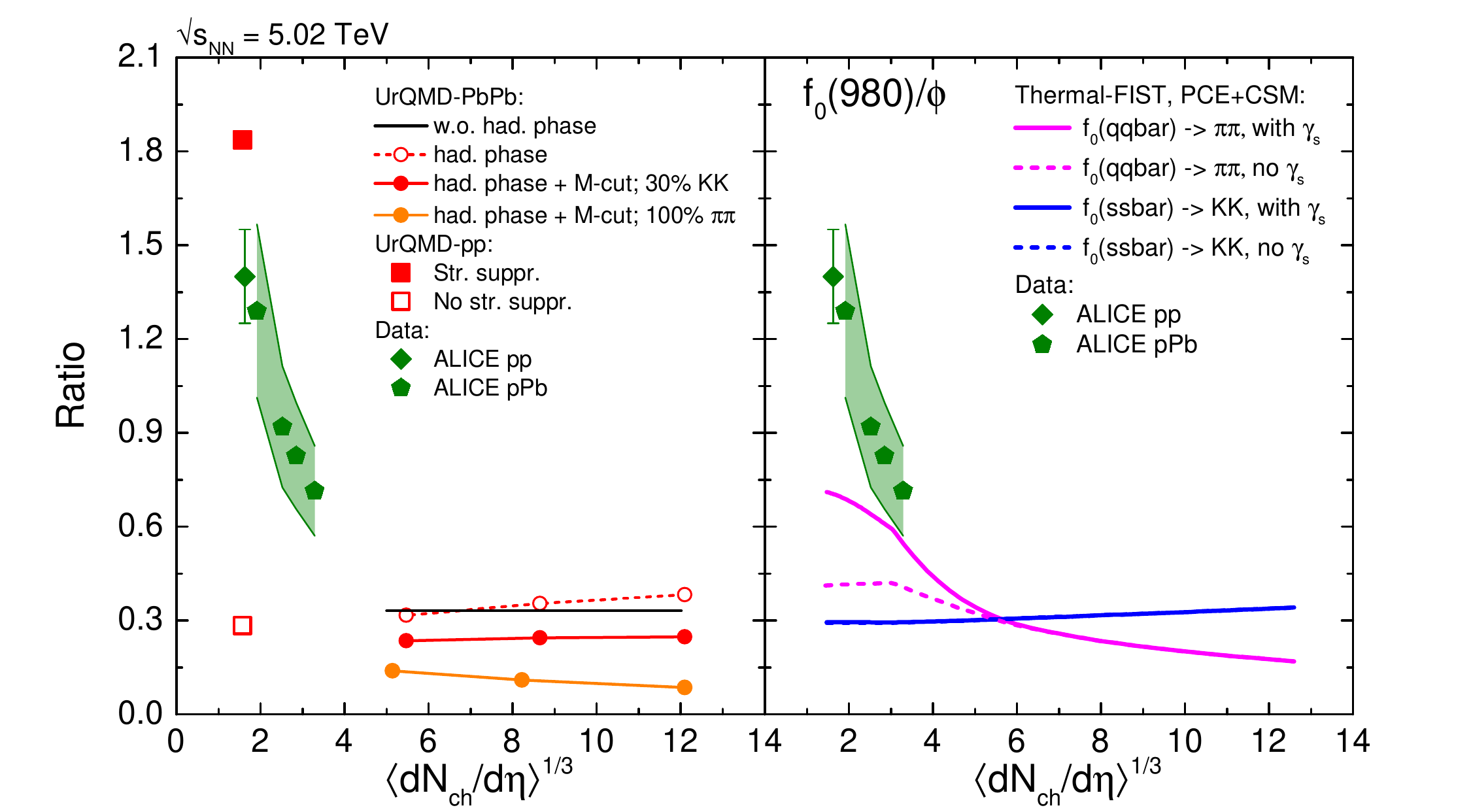}
  \caption{(Color online) The centrality dependence of the $f_0(980)/\phi$ ratio compared to ALICE data. The left panel compares the UrQMD calculations to the data while in the right panel the Thermal-FIST calculations are compared to the data. The UrQMD calculations are shown for p+p collisions with strangeness suppression (full red squares) and without strangeness suppression (open red squares). Also shown are the hybrid UrQMD calculations for Pb+Pb collisions without hadronic phase (solid black line) and with hadronic phase (open red circles, dashed line) with an additional phase space cut at 0.8 GeV with a 30\% branching ratio into kaons (full red circles) and at 0.8 GeV with a 100\% branching ratio into pions (full orange circles). The Thermal-FIST PCE+CSM calculations are shown considering the $f_0(980)$ as an $\langle\Bar{q}q\rangle$ state (magenta lines) with $\gamma_s$ factor (solid line) and without $\gamma_s$ factor (dashed line) and considering the $f_0(980)$ as an $\langle\Bar{s}s\rangle$ state (blue lines) with $\gamma_s$ factor (solid line) and without $\gamma_s$ factor (dashed line). Experimental data from the ALICE experiment at $\sqrt{s_{\mathrm{NN}}}= 5.02$ TeV is shown by green symbols \cite{ALICE:2022qnb,ALICE:2023cxn}.}
  \label{fig:f0_to_phi}
\end{figure*}

\subsection{The $f_0(980)$ to pion ratio}
Having discussed the role of the hadronic phase and system size dependence of resonance yields in general, and observational constraints for the $f_0(980)$ in particular, we can continue and study the centrality dependence of its ratios to different hadrons. Fig. \ref{fig:f0_to_pi} shows the centrality dependence of the $f_0(980)/\pi$ ratio compared to ALICE data \cite{ALICE:2022qnb}. The left panel shows the UrQMD calculations while in the right panel the Thermal-FIST calculations are compared to the data. Overall both the thermal and the UrQMD ratios underestimate significantly the ALICE p+p data and in the following we will discuss the systematic behavior as function of centrality. 

The UrQMD calculations are shown for p+p collisions (full red squares) and for Pb+Pb collisions in hybrid mode without hadronic phase (solid black line) and with a hadronic phase (colored lines with symbols). As expected, the ratio without a hadronic phase is independent of centrality since we use a fixed temperature of $T_{CF}=162$ MeV. The red dashed line shows results which include the hadronic phase, but no mass cut for the reconstruction, and use the conventional branching fraction of $\Sigma_{f_0\rightarrow \overline{K}K}/\Sigma_{tot}=30\%$. In this scenario we also observe an enhancement of the ratio towards more central collisions, due to the significant regeneration at low masses. Once the experimental mass cut for the $f_0(980)$ is applied, the ratio shows a suppression towards central collisions (red solid line with symbols) as most reconstructed $f_0(980)$ have lower masses. If we only allow the $f_0(980)$ to decay into pions and thus increase the regeneration through pions at low masses even further, the drop is significantly stronger (orange solid line with symbols) and essentially no measurable $f_0(980)$ are from regenerations.  

The Thermal-FIST PCE+CSM calculations, in the right panel, are shown considering the $f_0(980)$ as an $\langle\Bar{q}q\rangle$ state (magenta lines) and as an $\langle\Bar{s}s\rangle$ state (blue lines). The difference here is twofold. Firstly the $\langle\Bar{s}s\rangle$ state can be additionally suppressed by a strangeness suppression factor $\gamma_s$, which is seen in the difference between the solid and dashed lines\footnote{Note that the $\langle\Bar{q}q\rangle$ state relative to pions is slightly enhanced by including $\gamma_s$. This is an indirect effect from the modified pion yield obtained with $\gamma_s$ due to feeddown.}. 
As can be seen the strangeness suppression is only relevant for very small systems. 
The main difference, in the PCE fit, is related to the $f_0(980)$ decays, fixed to their quark content. In the fit, the $\langle\Bar{q}q\rangle$ state (magenta lines) is allowed to decay only into pions, while the $\langle\Bar{s}s\rangle$ state (blue lines) is allowed to decay only to kaons. As a result the 'kaon dominated' state is coupled to an overpopulation of kaons which allows for a stronger regeneration and thus higher final yield.

As demonstrated, both models show a clear sensitivity to the final multiplicity to the $f_0(980)$ branching ratios to kaons and pions.

\subsection{The $f_0(980)$ to $\phi$ ratio}

In the previous section we have discussed the ratio of $f_0(980)/\pi$, i.e. compared to the most abundant light quark state. In order to possibly understand the structure of the $f_0(980)$ a contrast with the well known $\phi$ mesons, a  $\langle\Bar{s}s\rangle$ state with similar mass, can be instructive.
In a naive thermal model the ratio of $f_0(980)/\phi$ should be approximately 1/3: due to their similar masses, the difference mainly comes from the spin degeneracies, $(2s_{f_0}+1)=1$ versus $(2s_\phi + 1) = 3$.

Figure \ref{fig:f0_to_phi} shows the centrality dependence of the $f_0(980)/\phi$ ratio of our two models compared to ALICE data \cite{ALICE:2022qnb,ALICE:2023cxn}. The left panel compares the UrQMD calculations to the data while in the right panel shows the Thermal-FIST PCE+CSM calculations.

Compared with the UrQMD model, an additional feature for p+p collisions has been added. Here, we compare p+p simulations with the standard strangeness suppression in the string break (full red squares) and without any strangeness suppression, i.e. light and strange quarks can be produced with the same probability (open red squares). In the string model the $f_0(980)$ is considered a pure light quark state and the $\phi$ a pure strange quark state. This comparison is included to highlight the effect of local strangeness suppression in the string which is very large, in fact a factor of 6 difference in the $f_0(980)/\phi$ is observed. One should keep in mind that this is so large due to the above-discussed feature of the string, which requires two pairs of $\langle\Bar{s}s\rangle$ to be produced adjacently to create a $\phi$. Most importantly, if no additional strangeness suppression of the $\phi$, as compared to the $f_0(980)$ is included, the ratio obtained is consistent with the naive thermal ratio of $1/3$. Therefore, a significant increase over this estimate would suggest that  $f_0(980)$ and $\phi$ have significantly different quark content.

Also shown in the left panel are the hybrid UrQMD calculations for Pb+Pb collisions without a hadronic phase (solid black line) and with a hadronic phase (colored lines with symbols).
The different scenarios are the same as in the $f_0(980)/\pi$ ratio and the conclusions are the same. A clear sensitivity to the final multiplicity to the $f_0(980)$ branching ratios to Kaons and pions is observed.

This conclusion is again supported by the Thermal-FIST PCE+CSM calculations shown in the right panel of figure \ref{fig:f0_to_phi}. If the $f_0(980)$ is considered a $\langle\Bar{s}s\rangle$ state (blue lines), the ratio to the $\phi$ is practically independent of centrality and the inclusion of any additional strangeness suppression and $f_0(980)/\phi \approx 1/3$. In the scenario where the $f_0(980)$ is a light quark state, a significant enhancement of the ratio for very small systems is observed, as expected, due to the suppression of the $\phi$. In addition, if the $f_0(980)$ has a dominant decay into two pions, suppression of the ratio towards central collisions is expected due to hadronic phase.

\section{Discussion}

It was shown that a measurement of the $f_0(980)$ resonance multiplicity, as a function of the system size, in ultra-relativistic collisions at the LHC can help pin down the internal structure of this elusive hadronic state. To illustrate this, we have employed the hybrid UrQMD model with a hadronic rescattering and the canonical statistical model with partial chemical equilibrium through Thermal-FIST to study the centrality dependence of various resonance to hadron ratios at the LHC with a special focus on the $f_0(980)$ state.

Three main conclusions could be drawn: 
\begin{enumerate}
\item Already by changing the $f_0(980) \rightarrow \bar{K}K$ partial branching from 0\% to 30\% the regeneration in the experimentally measured mass range changes significantly and thus a strong centrality dependence is observed. This allows us to pin down the relative branching fractions of the $f_0(980)$ which is important to determine the absolute $f_0$ yield from the observed decays.
\item A measurement of the $f_0(980)$ in the neutral $\pi^0\pi^0$ channel will allow us to investigate the full spectral function of the resonance.
\item The naive thermal ratio of $f_0(980)/\phi$ is 1/3 and driven by the corresponding ratio of spin degeneracies, given that their masses are nearly degenerate. 

From recent ALICE data a ratio of around 1 for p+p collisions is observed indicating a significant suppression of the $\phi$, by a factor of 3, as compared to the $f_0(980)$. This is interesting as it may be direct evidence that the $\phi$ is essentially $|\bar{s}s\rangle$ but the $f_0(980)$ is $|\Bar{q}q\rangle$ which leaves only the $\phi$ suppressed in small systems. 
\end{enumerate}

In general, the estimated yield of observable $f_0(980)$ resonances is significantly lower in all models compared to the ALICE pp data, i.e. both models underestimate the $f_0/\pi$ ratio in pp collisions, although the errors are large. As mentioned above, an important uncertainty is the branching fraction into the two pion channel which may be significant. Furthermore, particle production in pp collisions is not necessarily perfectly described in either the UrQMD or the Thermal-FIST model. Our predictions are more reliable for larger pPb or PbPb collisions. A measurement of the $f_0$ in heavy-ion reactions would therefore immediately shed some light on the uncertainties in the pp results.
This and the above conclusions make additional measurements of the $f_0$ in peripheral Pb+Pb collisions with ALICE, as well as future measurements at lower beam energies, for instance with the CBM experiment, necessary and worthwhile.

\begin{acknowledgments}
T.R. acknowledges support through the Main-Campus-Doctus fellowship provided by the Stiftung Polytechnische Gesellschaft Frankfurt am Main (SPTG). 
T.R. and J.S. thank the Samson AG for their support.
The authors acknowledge for the support the European Union's Horizon 2020 research and innovation program under grant agreement No 824093 (STRONG-2020). This research has received funding support from the NSRF via the Program Management Unit for Human Resources and Institutional Development, Research and Innovation [Grant No. B16F640076]. This work was supported by the DAAD (PPP Thailand).
The computational resources for this project were provided by the Center for Scientific Computing of the GU Frankfurt and the Goethe-HLR.
\end{acknowledgments}

\bibliography{refs}


\end{document}